\documentclass[journal]{IEEEtran}
\usepackage{cite}

  \usepackage{graphicx}
\usepackage[cmex10]{amsmath}
\ifCLASSOPTIONcompsoc
  \usepackage[caption=false,font=normalsize,labelfont=sf,textfont=sf]{subfig}
\else
  \usepackage[caption=false,font=footnotesize]{subfig}
\fi
\usepackage{fixltx2e}

\newcommand{\C}{{\rm I}\!\!\!{\rm C}}

\begin{document}
%
\title{Sliding window and compressive sensing for low-field dynamic magnetic resonance
imaging}
%
%
%

\author{Cristian Toraci, Gabriele Zaccaria, Stefano Ceriani, David Wilson,
Marco Fato, and Michele Piana%
\thanks{C. Toraci is with the IRCCS San Martino IST, Genova, Italy and the
Dipartimento di Matematica, Universit\`a di Genova, Genova, Italy}%
\thanks{G. Zaccaria is with the Dipartimento di Matematica, Universit\`a di
Genova, Genova, Italy}%
\thanks{M. Piana is with the Dipartimento di Matematica, Universit\`a di
Genova, Genova, Italy and CNR - SPIN, Genova}%
\thanks{D. Wilson is with the Department of Orthopaedics and the Centre for Hip Health and Mobility at the University of British Columbia and Vancouver Coastal Health Research Centre}%
\thanks{M. Fato and S. Ceriani are with the Dipartimento di Informatica,
Bioingegneria, Robotica e Ingegneria dei Sistemi, Universit\`a di Genova,
Genova, Italy}}

\markboth{Journal of \LaTeX\ Class Files,~Vol.~11, No.~4, December~2012}%
{Shell \MakeLowercase{\textit{et al.}}: Bare Demo of IEEEtran.cls for Journals}
%



\maketitle

\begin{abstract}
We describe an acquisition/processing procedure for image reconstruction in dynamic Magnetic Resonance Imaging (MRI). The approach requires sliding window to record a set of trajectories in the k-space, standard regularization to reconstruct an estimate of the object and compressed sensing to recover image residuals. We validated this approach in the case of specific simulated experiments and, in the case of real measurements, we showed that the procedure is reliable even in the case of data acquired by means of a low-field scanner.
\end{abstract}


%
\IEEEpeerreviewmaketitle

\section{Introduction}

Magnetic resonance imaging (MRI) is a tomographic technique that produces
images of some internal chemical characteristic of an object from induced
nuclear magnetic resonance signals. Many clinical applications in MRI involve
imaging of non-static anatomies, like in the case of abdominal and cardiac
imaging, while other applications can take advantages from dynamic imaging of
objects in motion, like in the case of studies of joint kinematics.

The signal generated by an MRI system is the Fourier transform of the image of
interest, sampled on a finite portion of the spatial frequencies using a
limited number of points. The region that must be acquired to reconstruct a
given image with the desired spatial resolution is usually called
{\em{k-space}}: k-space data are acquired along continuous trajectories whose
length is limited by sampling bandwidth and signal decay due to relaxation. The
acquisition time of a whole MRI exam is roughly proportional to the number of
sampled trajectories. Therefore, in dynamic imaging, this number should be
optimized to realize a trade-off between undersampling and motion blur
artifacts.  To address this issue several strategies have been proposed:

\begin{itemize}
\item Historically, Echo Planar Imaging
(EPI)~\cite{zigzag_epi,epi,multishot_epi} allows full coverage of the k-space
in a very short time (40 to 100 ms) thus reducing motion-related artifacts. Its
main drawbacks are the use of demanding hardware characteristics and notable
geometric distortions induced by inhomogeneities of the static magnetic
field.
\item Parallel imaging~\cite{pmri_rev1,pmri_rev2} utilizes phase-array coils
whose receiving elements have a characteristic space-dependent sensitivity
distribution. The redundancy given by multiple elements can be used to
reconstruct under-sampled data. In parallel imaging, however, there are
limitations due to the kind of acquisition protocol and hardware adopted such
as signal-to-noise (SNR) ratio, image geometry or maximal achievable
acceleration factors.
\item A more general approach comes from constrained reconstruction techniques
that aim at integrating partial k-space data using \emph{a-priori}
information~\cite{prior_information}. These might be a high resolution
reference image, acquired before or after the fast acquisition, or some known
property of the image, like sparsity in a proper basis.
\end{itemize}

In this last framework, compressed
sensing~\cite{compressed_sensing, candes_romb_tao}
provides computational tools able to reconstruct sparse signals from an
under-sampled set of measurements. Although MRI images are not in general
sparse, sparsifying transforms can be introduced~\cite{sparse_mri, dynamic_cs}.
For example, kt-FOCUSS~\cite{kt_focuss} records one or two fully sampled data
sets and several under-sampled data sets. Then RIGR~\cite{rigr} or motion
estimation and compensation techniques used in video encoding~\cite{mpeg} are
applied to compute initial {\em{estimates}} for all the time samples; finally,
sparsification is used for reconstructing the difference ({\em{residuals}})
between optimal images and estimates. The main drawback of kt-FOCUSS is in the
fact that recording high-resolution volumes requires a long acquisition time
and thus notably limits the applicability of the method in the clinical
workflow. 

The goal of the present paper is to apply compressed sensing  in order to
reconstruct dynamic MRI volumes without the time-demanding acquisition of high
resolution data. An example of this approach is given in~\cite{sw_tv}, where
sparsification is achieved by differentiating along the temporal direction. We
observe that numerical differentiation is numerically unstable. This drawback
is not an issue for~\cite{sw_tv}, where data acquired by means of a
high-field MRI scanner are used. Here we are interested in applications at
low-field data and in this case the low signal-to-noise ratio may impact on the
reconstruction quality. Therefore we realize this approach by means of a
different sparsification scheme and show its effectiveness in the case of data
at $0.5$ T. More specifically, in such scheme:
\begin{enumerate}
\item A sliding window  approach~\cite{pulse_handbook} records a set of trajectories of the body in k-space.
\item\label{item:strategy-estimate} A Fourier-based inversion technique
provides a first estimate of the object, which is affected by motion blur, by
utilizing a high enough number of trajectories.
\item\label{item:strategy-residual} A kt-FOCUSS-like approach uses this blurred
estimate and compressed sensing to reconstruct the residuals. 
\item Summing up the residuals and the estimates provides a set of images, each
one corresponding to a very limited number of trajectories (and therefore to a
very limited temporal range in the acquisition time).
\end{enumerate}

We observe that, in item \ref{item:strategy-estimate}), the computation of the
estimate uses fully or nearly-fully sampled data and can be realized by using
standard regularization methods for image reconstruction. On the other hand,
item \ref{item:strategy-residual}) relies on the fact that de-blurring can be
realized using compressed sensing as shown in~\cite{cs_motion_correction}. As
far as the implementation of the compressed sensing step is concerned, we
observe that the identification of the optimal technique in this field is still
an open research issue and therefore we decided to compare the performances
provided by two different methods: Orthogonal Matching Pursuit
(OMP)~\cite{omp}, that recovers the sparse
components of the image in a \emph{greedy} fashion, thus limiting the
computational effort of the processing; and an optimization method with
$\ell_1$ penalty term where the minimum problem is solved by using a
split-Bregman algorithm.

The content of the paper is organized as follows. In Section \ref{sec:model} we
describe the imaging model and the reconstruction strategy. Section
\ref{sec:implementation} contains the details of the method implementation.
Section \ref{sec:examples} contains some numerical results using synthetic data
simulated according to realistic procedures and Section \ref{sec:cervical}
shows how this procedure works in the case of a set of experimental
measurements of the cervical spine.  Our conclusions are offered in Section
\ref{sec:conclusion}.

\section{Model and reconstruction strategy}
\label{sec:model}

Data formation in MRI can be mathematically modeled by means of a Fourier
transform relation. More precisely, at a given acquisition time, MRI records
samples of the Fourier transform of the function of interest along trajectories in the k-space. At the
end of a sequence of acquisition times, the data is represented by a set of
trajectories. More formally, let $x \in \C^{N \times N}$ be a discretized
version of the function representing the object of interest (i.e., $x$ is the
unknown in the image space) and $y \in \C^{M \times K}$ is the measured data
(i.e., in the k-space, a set of $M$ trajectories made of $K$ sampled frequencies): then, using a lexicographic ordering, $y$ and $x$ are related
by the matrix equation
\begin{equation}\label{model}
y = Fx,
\end{equation}
where $F \in \C^{M K \times N^2}$ is a linear transformation obtained from a
(non-uniform) discretization of the Fourier transform. 

The MRI reconstruction problem is the one to determine an estimate of $x$ given
$y$. In the case of static imaging, if the number of trajectories $M$ is high
enough and the criterion with which they sample the k-space is Nyquist-like,
this image reconstruction problem is well-posed and inverse Fourier
transform provides reliable results.  However, full
(or high) k-space coverage requires a long acquisition time which implies, for
dynamic studies, that images obtained by means of Fourier-based methods are
affected by motion blur. On the other hand, motion blur can be reduced by using
fewer trajectories but in this case other artifacts are introduced, that are due
to the limited number of Fourier components one utilizes for the inversion. A
computational strategy based on compressed sensing may help to overtake this
deadlock. Let us denote with $y_{\nu}$
the set of $\nu$ trajectories acquired in a time interval $\delta t$ around
time $t$. Then the reconstruction $x_{\nu}$ corresponding to $y_{\nu}$ can be
written as
\begin{equation}
\label{sum}
x_{\nu}=x_{M} + \delta x,
\end{equation}
where $x_{M}$ is the reconstruction obtained by applying some regularization
method to the data $y_{M}$ corresponding to the complete set of $M$
trajectories. If $F_{\nu}$ is the linear transformation producing the $\nu$
trajectories from a given input data, we have that

\begin{equation}
\label{F-sum}
F_{\nu} x_{\nu} = F_{\nu} x_{M} + F_{\nu} \delta x.
\end{equation}

Assuming that $\delta x$ is sufficiently sparse justifies the application of
the constrained minimum problem

\begin{equation}
\label{compressed-sensing}
\left\{ \begin{array}{l}
\min \|\delta x\|_1 \\
\|F_{\nu} \delta x - (y_{\nu} - F_{\nu} x_M) \|_2^2 < \epsilon,
\end{array}
\right.
\end{equation}
with $\epsilon$ sufficiently small. In order to implement such reconstruction
strategy, three issues must be addressed:

\begin{itemize}
\item The data sets $y_{M}$ and $y_{\nu}$ must have specific properties: $M$
must be sufficiently big such that the corresponding estimate $x_M$ is not
affected by artifacts due to undersampling the k-space; further, $y_{\nu}$ must
be appropriate to compressed sensing, i.e. the corresponding $\delta x$ must be
sufficiently sparse and the undersampling must be lower near the center of k-space and higher near the periphery~\cite{sparse_mri}.
\item In equation (\ref{compressed-sensing}) both $\delta x $ and $x_M$ are
unknown. Therefore a regularization method must be applied in order to
reconstruct the estimate $x_M$.
\item Once $x_M$ has been reconstructed, a computational method is required to
solve the constrained minimum problem (\ref{compressed-sensing}) enhancing the
sparsity of the solution.
\end{itemize}

In the following section we provide details about the implementation of the
previous issues as adopted in the numerical applications.

\section{Implementation}
\label{sec:implementation}

The reconstruction paradigm formulated in this paper requires a specific
acquisition strategy, which is based on a sliding window scheme
\cite{view_sharing}.  Let $k_m: [0,\tau] \rightarrow [-\pi,\pi]^2$ be a
sampling trajectory in the two-dimensional k-space, such that $k_m(t)$ is the
parametric representation of the $m$-th trajectory, $t$ runs in the interval
$[0,\tau]$ and $\tau$ is the acquisition time specific for that trajectory.
Further, assume that the time required to sample a trajectory is short enough
to have negligible motion in this interval. Then choose a set of trajectories
$\{k_m\}_{m=1}^{M}$ in such a way that:
\begin{enumerate}
\item\label{item:swcs-hp-sw} The set $k_{m_1} \cap k_{m_2}$, $\forall m_1 \neq
m_2$; $m_1,m_2 = 1,\ldots,M$ is made of a small number of frequencies.
\item\label{item:swcs-hp-nyq} The sampling of the k-space given by
$\cup_{m=1}^M k_m$ is appropriate for solution of (\ref{model}) by means of
some regularization technique.
\item\label{item:swcs-hp-cs} For some integer $\nu$, $0<\nu < M$, the sampling
given by $\cup_{m=m_0-\nu/2}^{m_0 + \nu/2} k_m$ is appropriate for inversion by
means of a compressed sensing technique for all $m_0$ such that $1 + \nu/2 \leq
m_0 \leq M - \nu/2$.
\end{enumerate}
Items \ref{item:swcs-hp-sw}) and \ref{item:swcs-hp-nyq}) imply that the k-space
is sampled in a complete way, so that the data $y_M$ corresponding to
$\cup_{m=1}^M k_m$ provide a good (but blurred) estimate $x_M$ by means of some
conventional inversion technique. Specifically, in our applications we utilized
a Conjugate Gradient scheme to solve the weighted least-squares problem
\begin{equation} \label{tikhonov} x_M = \arg \min_x \|F^W x - y^W_M\|_2^2
\end{equation}
applied against  a windowed version $y^W_M$ of the data $y_M$. The windowed
data is defined as follows: if $ y_m$ represents the data from trajectory $k_m
$, then $ y_m^W = y_m h_M(m) $, with $h_M$ the Hamming window defined as $
h_M(m) = 0.54 - 0.46 \cos ( 2 \pi m / M ) $. Analogously, $F^W = W F$ where $W$
is a diagonal matrix whose non-zero entries are given by $h_M(m)$ for
$m=1,\ldots,M$.


On the other hand, item \ref{item:swcs-hp-cs}) is not specific of sliding
window but it is necessary for a compressed sensing approach to the
reconstruction of the residual $\delta x$. In fact, once $x_M$ is determined by
(\ref{tikhonov}), we computed $\delta x$ by solving (\ref{compressed-sensing})
with two methods: an Orthogonal Matching Pursuit (OMP) algorithm and an
$\ell_1$ optimization method. 

\subsection{Orthogonal Matching Pursuit}

OMP is an iterative scheme that implements sparsity enhancement according to a
\emph{greedy} fashion. The main idea is based on the observation that a signal
with $s$ non-zero components generates data which are the linear combination
of just $s$ columns of the matrix modeling the data formation. The task of OMP
is to iteratively select the column which is mostly correlated with the data
and solve the reconstruction problem using just the selected columns.

%

In order to improve the speed of the method, in the numerical applications
described in the next section we applied a modification of OMP, named KOMP
(i.e., $K$-fold Orthogonal Matching Pursuit)~\cite{komp} whereby, each
iteration step chooses the $K>1$ most correlated columns.
%
%
It can be proved that this modification has the same convergence properties of
OMP~\cite{romp, cosamp}. At each iteration step, the reconstruction problem
limited to the selected columns is solved by means of a Conjugate Gradient
algorithm.

\subsection{$\ell_1$ optimization}

Equation (\ref{compressed-sensing}) can be determined by recasting it as the minimum problem
\begin{equation}\label{compressed-tikhonov}
\delta x = \arg \min_{\delta x} \|F_{\nu} \delta x - (y_{\nu} - F_{\nu} x_M)
\|^2_2 + \lambda \| \delta x \|_1
\end{equation}
where $\lambda$ plays the role of a regularization parameter. In our
applications this problem is solved by using the
split-Bregman  method~\cite{split_bregman} which transforms equation
(\ref{compressed-tikhonov}) into the new minimum problem (in $\delta x$ and the
slack variable $u$)
\begin{align}
\label{split-Bregman}
(\delta x, u) &= \arg \min_{\delta x, u} \|F_{\nu} \delta x - (y_{\nu} -
F_{\nu} x_M) \|^2_2 \nonumber \\ &+ \lambda_1 \|u - \delta x\|^2_2 + \lambda_2
\| u \|_1
\end{align}
This equation is iteratively solved for $\delta x$ (with fixed $u$) by means of
Conjugate Gradient and analytically by soft threshold for $u$ (with fixed
$\delta x$). At each iteration the data is updated by adding the data residual
computed by solving the forward problem.

In both KOMP and the split-Bregman-based approach some regularization parameters must be optimally fixed (the integer number $K$ for KOMP and the regularization parameters $\lambda_1$ and $\lambda_2$ for $\ell_1$ minimization). In the simulated experiments, we applied an optimality criterion, i.e. we minimized the root mean square error between the parameterized reconstructions and the ground truth. In the application against a set of experimental data, we chose for these parameters values of the same order of magnitude as the ones obtained in the simulations and exploited the notable robustness of the reconstruction with respect to variation in this parameters.


\section{Numerical examples}
\label{sec:examples}

We considered two numerical tests for the validation of this sliding window / compressed sensing approach (SWCS from now on) to MRI. The first test aims at determining the amount of spatial resolution that can be recovered thanks to compressed sensing. The second test  is to evaluate the overall accuracy of the imaging process in the case of a digital phantom. 

\subsection{Test 1: spatial resolution}

Our imaging model is made of two identical 2D Gaussian functions with same
standard deviation $\sigma$, and centered at
\begin{equation}\label{center}
\mu_y = 0~~~~~\mu_x = \pm v \cdot t,
\end{equation}
where $(0,0)$ is the center of the image. We consider three different velocity
values equal to $0.032$, $0.064$ and $0.128$ pixel per frame, with $1000$
frames $t \in [-500,500]$ and five values of $\sigma = 2, 4, 6, 8, 10$. The
signal was sampled using radial trajectories separated by the \emph{golden
angle}~\cite{golden_angle}. Each trajectory, corresponding to one time point,
sampled 512 k-space points at twice the Nyquist ratio. Images were
reconstructed using 305 projections for estimates and 72 projections for
residuals.

\begin{figure}[!t]
\centering
\subfloat{\includegraphics[width=8cm]{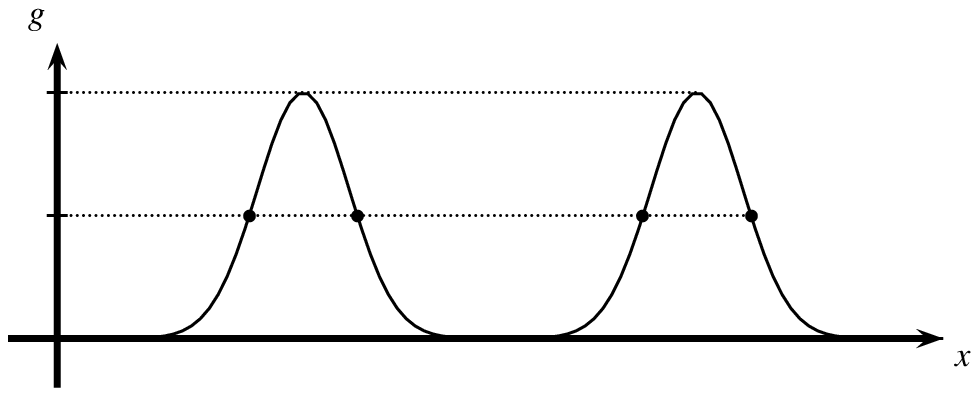}%
\label{fig:gaussian-profiles-sep}}
\hfil
\subfloat{\includegraphics[width=8cm]{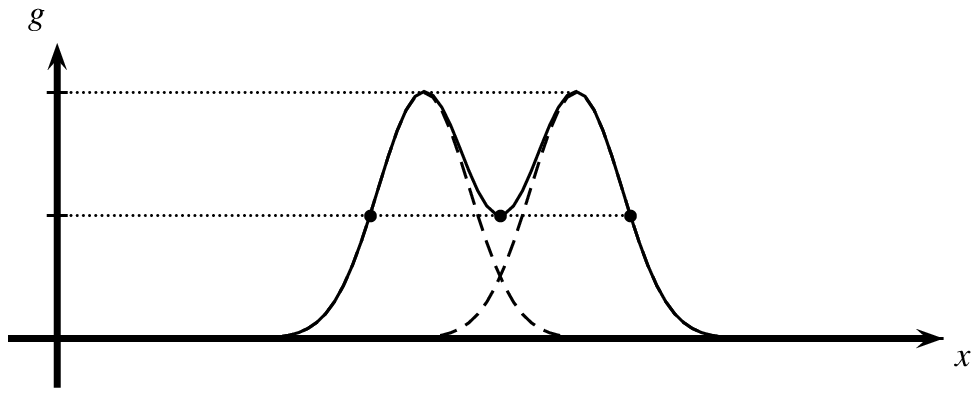}%
\label{fig:gaussian-profiles-mix}}
\hfil
\subfloat{\includegraphics[width=8cm]{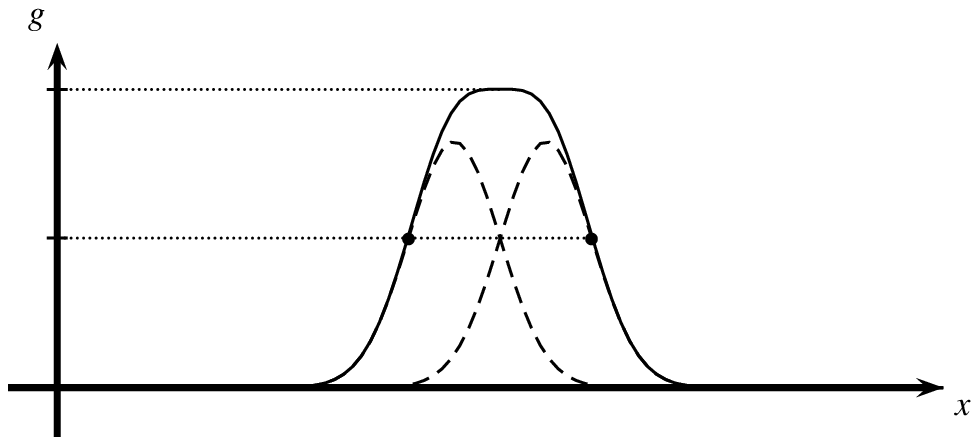}%
\label{fig:gaussian-profiles-sup}}
\caption{Gaussians profiles at three key time points:
\ref{fig:gaussian-profiles-sep} fully separated Gaussians at $ t = 0 $;
\ref{fig:gaussian-profiles-mix} unresolved Gaussians at $ t = t_0 $ and $ t =
t_1 $; \ref{fig:gaussian-profiles-sup} Fully superimposed Gaussians at $ t =
t_2 $ and $ t = t_3 $. In \ref{fig:gaussian-profiles-mix} and
\ref{fig:gaussian-profiles-sup} The dashed line represents the individual
Gaussians, the solid line is the sum of the two}
\label{fig:2gaussians}
\end{figure}

\begin{figure}[!t]
\centering
\includegraphics[width=8cm]{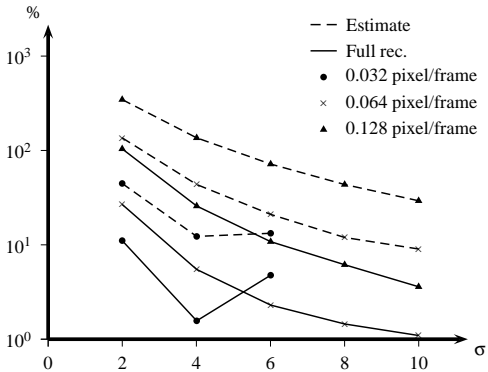}
\caption{Measures of FWHM of the Gaussians in estimates and fully reconstructed
images. The plot represents relative percentage errors between theoretical
values and measures. Data for $ \alpha = 0.032$~pixel/frame at $\sigma =$ 8 and
10 pixels are not available, since Gaussians are not resolved at any time frame
in these experiments.}
\label{fig:fwhm}
\end{figure}

\begin{figure*}[!t]
\centering
\includegraphics[width=18cm]{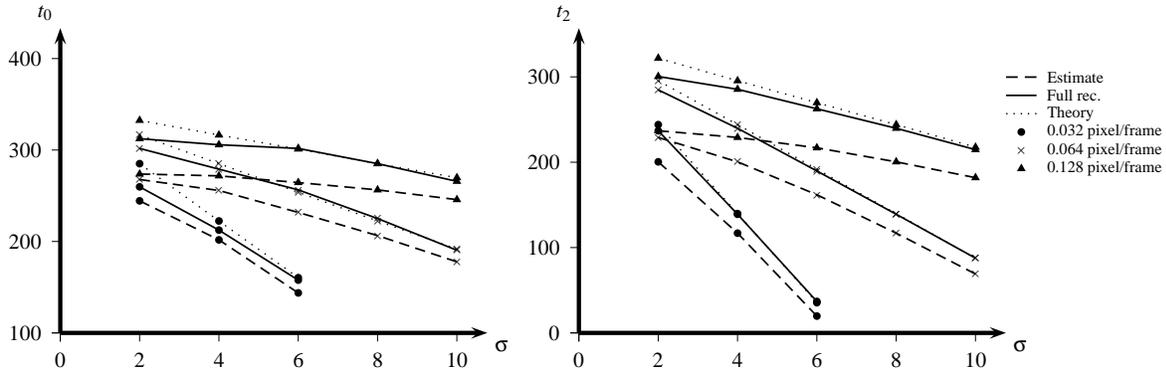}
\caption{Measures of $ t_0 $ (left) and $ t_2 $ (right) for estimates and full
reconstructions, in comparison with theoretical values. Data for $ v =
0.032$~pixel/frame at $\sigma =$ 8 and 10 pixels are not available, since
Gaussians are fully superimposed at any time frame in these experiments.}
\label{fig:times}
\end{figure*}

In the analysis of the reconstructed images we measured (see Figure
\ref{fig:2gaussians}):
\begin{itemize}
\item The Full Width at Half Maximum (FWHM) at the first reconstructed frame.
\item The frame $t_0$ at which the grey level at the mid point between the {\em{approaching}} two Gaussian functions is equal to half maximum.
\item The frame $t_1$ at which the grey level at the mid point between the two {\em{parting}} Gaussian functions is equal to half maximum.
\item The last frame $t_2<0$ at which the two Gaussian peaks can be distinguished.
\item The first frame $t_3 >0$ at which the two Gaussian peaks can be distinguished again.
\end{itemize}
The FWHM values for all $\sigma$ values are in Figure \ref{fig:fwhm}; the values of $t_0$ and $t_2$ are in Figure \ref{fig:times}. As expected, the values of $t_1$ and $t_3$ are essentially symmetrical with respect to $t_0$ and $t_2$ respectively, and are not shown. The figures are obtained by using $\ell_1$-optimization but the results provided by KOMP are analogous: in both cases the use of compressed sensing improves the spatial resolution of the reconstruction (and, accordingly, times $t_0$ and $t_2$ decrease). As expected, this effect is more significant for higher values of the speed, i.e. when motion blur is more intense.


\subsection{Test 2: image quality}

\begin{figure}[!t]
\centering
\includegraphics[width=9cm]{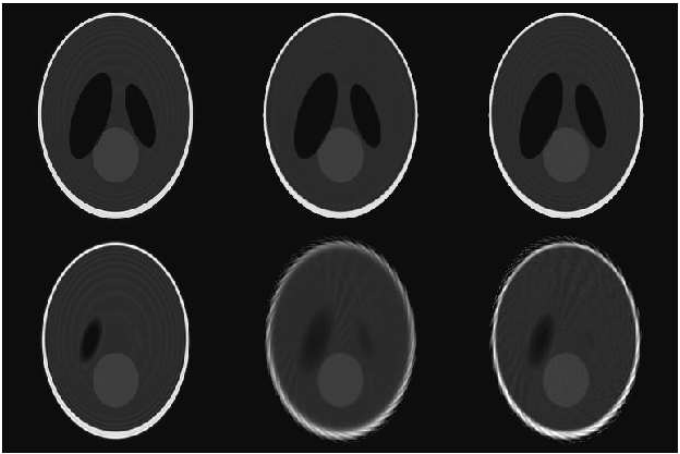}
\caption{Shepp-Logan images for time frame number 300 for two different speeds.
Top row is speed $p = 0.01$; bottom row is speed $p = 0.02$. Left column is
ground truth images; central column is estimate imagess; and Right column is
the final reconstruction, sum of estimates and CS residuals. While on lower
speed the contribution of CS is minimal, due to low blurring levels, resolution
improvements are evident at higher speed. Morevoer at higher speed the left
ventricle is slightly visible on estimates, but is clearly an artifact; on the
final images this artifact disappears.}
\label{fig:shepp-logan}
\end{figure}

For this experiment we considered the Shepp-Logan three-dimensional digital
phantom and simulated one-dimensional motion along the direction orthogonal to
the image plane. The pixel displacement is 
\begin{equation}\label{displacement}
d = \Delta x \cdot p,
\end{equation}
where $\Delta z$ is the slice thickness and the parameter $p$ mimics the motion
velocity. In our applications we assumed $p=0.01, 0.02, 0.03, 0.04$. To avoid
inverse crime, data were generated directly in k-space, using the analytical
formula for the Fourier transform of the piecewise function modeling the
phantom~\cite{kspace_phantom,3d_kspace_sl}. We used the same trajectories as in
the previous example and time sampling has been chosen in such a way the at
$t=0$ the imaging plane is at the center of the phantom. The measurements in
the k-space were affected by Gaussian noise with standard deviation $\sigma = 5
\times 10^{-4}$ times the signal maximum. The results of this experiment are in
Table \ref{table:shepp-logan-tab}, containing the root mean square error (RMSE) for
the blurred estimate and for SWCS with KOMP and the split-Bregman algorithm, corresponding to two
specific time points. Further, Figure \ref{fig:shepp-logan} visually
illustrates the reconstructions for just two values of $p$, namely $ p=0.01,
0.04 $. Both the table and the figure show that the SWCS improves the quality of the reconstruction, particularly for
high values of $p$ while the improvement is negligible at low speed, when the
correction demand is low. We point out that in both the images in the figure
and the computation of the reconstruction errors we applied a binary mask that
excludes the background.

\begin{table}
\caption{RMSE for two sample frames of the Shepp-Logan experiment.}
\label{table:shepp-logan-tab}
\centering
\subfloat{%
\begin{tabular}{c||c|c|c|c}
\hline
152th time frame & p = 0.01 & p = 0.02 & p = 0.03 & p = 0.04 \\
\hline
\hline
estimate         & 0.084    & 0.15     & 0.22     & 0.32 \\
\hline
KOMP             & 0.083    & 0.13     & 0.16     & 0.21 \\
\hline
Split-Bregman    & 0.083    & 0.13     & 0.16     & 0.21 \\
\hline
\end{tabular}%
}
\vfil
\subfloat{%
\begin{tabular}{c||c|c|c|c}
\hline
600th time frame & p = 0.01 & p = 0.02 & p = 0.03 & p = 0.04 \\
\hline
\hline
Estimate         & 0.088    & 0.16     & 0.25     & 0.39 \\
\hline
KOMP             & 0.089    & 0.14     & 0.16     & 0.24 \\
\hline
Split-Bregman    & 0.088    & 0.14     & 0.14     & 0.24 \\
\hline
\end{tabular}%
}
\end{table}

\section{Application to orthopedic imaging}
\label{sec:cervical}

\begin{figure}[!t]
\centering
\includegraphics[width=9cm]{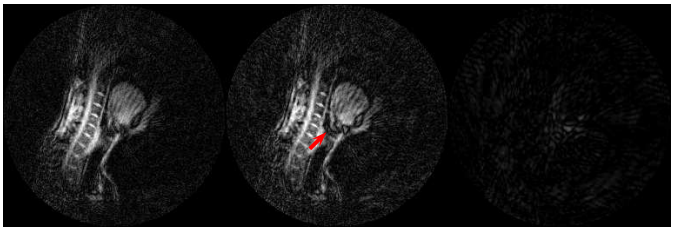}
\caption{Sample frame of the volunteer scan. The image on the left is the estimate; the central
image is the full reconstruction; the image on the right is the residual. The most
significant improvement can be seen in the region below the tongue, marked with
the arrow.}
\label{fig:real}
\end{figure}

We scanned a human subject using a 0.5~T MRI system with an open U-shaped
magnet (Mr Open, Paramed Medical System, Genova, Italy). We performed cervical spine
imaging using a two-channel dedicated coil according to the following
parameters: TE/TR 7.3/20~ms; FA 30$^\mathrm{o}$; sagittal plane; FOV 300~mm;
512~samples; sampling bandwidth $\pm$50~kHz; slice thickness 6~mm;
1000~time frames. With these parameters we achieved a time sampling of 50
frames per second. The subject was instructed to flex and extend his neck 
during the acquisition at 3 cycles per minute. Figure \ref{fig:real} presents
the estimate provided by conjugate gradient when applied against the whole data
set (corresponding to 305 trajectories), the residual restored by compressed
sensing when using just 72 trajectories (this is the implementation based on
$\ell_1$ optimization) and the sum of the two. This figure shows that,
particularly in the region below the tongue, SWCS is able to recover some high resolution details hidden in the
reconstruction provided by the conjugate gradient estimate.

\section{Conclusions}
\label{sec:conclusion}

The present paper describes a novel strategy for acquisition and reconstruction
of images from dynamic MRI data. The approach exploits a sliding window
paradigm for the recording of trajectories in k-space, a standard
regularization method for the reconstruction of an initial estimate
corresponding to a complete data set and compressed sensing techniques for the
reconstruction of the residual images in correspondence of narrow time
intervals. We showed that this strategy is reliable for improving both the
achievable spatial resolution and the overall reconstruction quality, even in the case of data acquired with low-field systems. 
The strategy does not significantly depend on the kind of sparsity enhancing
technique applied during the compressed sensing step. From a computational
viewpoint, we realized that, in the numerical tests performed, the optimal
values for $K$ in KOMP is rather high. Accordingly, the optimal value of the
regularization parameter controlling the $\ell_1$ penalty term in split-Bregman
is rather small. This suggests the need of a more systematic investigation of
the interplay between the roles of sparsity and sliding window acquisition in
the image reconstruction process. Finally, in order to systematically study the
applicability power of this approach it is now necessary to validate its
performances against experimental data acquired in clinical contexts.


%

%
%
%
\section*{Acknowledgment}
Cristian Toraci is supported by a PO CRO FSE 2007/13 Asse IV grant from Regione Liguria, Italy. The experimental part described in the paper has been performed in Vancouver, at the Center for Hip Health and Mobility and was supported by an operating grant from the Canadian Institutes of Health Research.

\ifCLASSOPTIONcaptionsoff
  \newpage
\fi



\bibliographystyle{IEEEtran}
\bibliography{shorttitles,ctoraci,zaccaria}
\end{document}